\documentclass[]{article}
\usepackage{lmodern}
\usepackage{amssymb,amsmath}
\usepackage{ifxetex,ifluatex}
\usepackage{fixltx2e} 
\ifnum 0\ifxetex 1\fi\ifluatex 1\fi=0 
  \usepackage[T1]{fontenc}
  \usepackage[utf8]{inputenc}
\else 
  \ifxetex
    \usepackage{mathspec}
  \else
    \usepackage{fontspec}
  \fi
  \defaultfontfeatures{Ligatures=TeX,Scale=MatchLowercase}
\fi
\IfFileExists{upquote.sty}{\usepackage{upquote}}{}
\IfFileExists{microtype.sty}{%
\usepackage{microtype}
\UseMicrotypeSet[protrusion]{basicmath} 
}{}
\usepackage[margin=1in]{geometry}
\usepackage{hyperref}
\hypersetup{unicode=true,
            pdftitle={Facilitating information system development with Panoramic view on data},
            pdfauthor={Dejan Lavbič, Iztok Lajovic, and Marjan Krisper},
            pdfborder={0 0 0},
            breaklinks=true}
\usepackage{natbib}
\usepackage{longtable,booktabs}
\usepackage{graphicx,grffile}
\makeatletter
\def\maxwidth{\ifdim\Gin@nat@width>\linewidth\linewidth\else\Gin@nat@width\fi}
\def\maxheight{\ifdim\Gin@nat@height>\textheight\textheight\else\Gin@nat@height\fi}
\makeatother
\setkeys{Gin}{width=\maxwidth,height=\maxheight,keepaspectratio}
\IfFileExists{parskip.sty}{%
\usepackage{parskip}
}{
\setlength{\parindent}{0pt}
\setlength{\parskip}{6pt plus 2pt minus 1pt}
}
\providecommand{\tightlist}{%
  \setlength{\itemsep}{0pt}\setlength{\parskip}{0pt}}
\ifx\paragraph\undefined\else
\let\oldparagraph\paragraph
\renewcommand{\paragraph}[1]{\oldparagraph{#1}\mbox{}}
\fi
\ifx\subparagraph\undefined\else
\let\oldsubparagraph\subparagraph
\renewcommand{\subparagraph}[1]{\oldsubparagraph{#1}\mbox{}}
\fi

\let\rmarkdownfootnote\footnote%
\def\footnote{\protect\rmarkdownfootnote}

\usepackage{titling}

  \title{Facilitating information system development with Panoramic view on data}
    \pretitle{\vspace{\droptitle}\centering\huge}
  \posttitle{\par}
    \author{Dejan Lavbič, Iztok Lajovic, and Marjan Krisper}
    \preauthor{\centering\large\emph}
  \postauthor{\par}
    \date{}
    \predate{}\postdate{}
  
\usepackage{booktabs}
\usepackage{longtable}
\usepackage{array}
\usepackage{multirow}
\usepackage[table]{xcolor}
\usepackage{wrapfig}
\usepackage{float}
\usepackage{colortbl}
\usepackage{pdflscape}
\usepackage{tabu}
\usepackage{threeparttable}
\usepackage{threeparttablex}
\usepackage[normalem]{ulem}
\usepackage{makecell}

\usepackage{amsthm}

\theoremstyle{definition}

\theoremstyle{definition}

\theoremstyle{definition}

\theoremstyle{remark}

\begin{document}
\maketitle

\begin{quote}
\textbf{Dejan Lavbič}, Iztok Lajovic, and Marjan Krisper. 2010.
\href{https://doi.org/10.2298/CSIS091122031L}{\textbf{Facilitating
information system development with Panoramic view on data}},
\href{http://www.comsis.org/}{Computer Science and Information Systems
\textbf{(ComSIS)}}, 4(7), pp.~737 - 767.
\end{quote}

\section*{Abstract}\label{abstract}
\addcontentsline{toc}{section}{Abstract}

The increasing amount of information and the absence of an effective
tool for assisting users with minimal technical knowledge lead us to use
associative thinking paradigm for implementation of a software solution
-- Panorama. In this study, we present object recognition process, based
on context + focus information visualization techniques, as a foundation
for realization of Panorama. We show that user can easily define data
vocabulary of selected domain that is furthermore used as the
application framework. The purpose of Panorama approach is to facilitate
software development of certain problem domains by shortening the
Software Development Life Cycle with minimizing the impact of
implementation, review and maintenance phase. Our approach is focused on
using and updating data vocabulary by users without extensive
programming skills. Panorama therefore facilitates traversing through
data by following associations where user does not need to be familiar
with the query language, the data structure and does not need to know
the problem domain fully. Our approach has been verified by detailed
comparison to existing approaches and in an experiment by implementing
selected use cases. The results confirmed that Panorama fits problem
domains with emphasis on data oriented rather than ones with process
oriented aspects. In such cases the development of selected problem
domains is shortened up to 25\%, where emphasis is mainly on analysis,
logical design and testing, while omitting physical design and
programming, which is performed automatically by Panorama tool.

\section*{Keywords}\label{keywords}
\addcontentsline{toc}{section}{Keywords}

Software development, associative thinking, object recognition, rapid
application development

\section{Introduction}\label{introduction}

There is a vast amount of data available on a daily basis from various
sources that we have to perceive, analyse and comprehend. Several
approaches to organising data on computers exist, starting from
straightforward support for simple data structures that majority of
operating systems offer, to sophisticated database management systems
for storage of composed and complex data structures
\citep{harbron_file_1988, tharp_file_1988}. Every approach deals with
management of data in its own distinct way, where user is able to enter
and edit the data as well as prepare and view reports. The key question
is how a user without extensive IT knowledge can successfully cope with
the complexity of all perceived data.

The development of Computer Science was always focused on aiding human
actors at performing tasks
\citep{caramia_mining_2006, tell_towards_1992, ortner_enterprise_1991, lavbic_ontology-based_2010}.
In this paper, Panorama as an implementation of our approach to support
human actors in information management and software development will be
discussed. Panorama is based on principles of associative thinking and
it supports inclusion of various data structures that compound our
information space and facilitates traversing through data by following
associations. This concept is based on the human observation of the
environment and its context -- the complete information about selected
object is available at certain time with a possibility of further
exploration of related concepts through associations. The present paper
presents Panorama approach with object recognition process, based on
context + focus information visualization techniques
\citep{ware_information_2004}, that starts with observation and
perception of the environment and continues with context of observed
object. This paper also addresses how software development is
facilitated with Panorama tool. There a user can develop, use and
maintain an information system that would suit his/her needs for
information management without an extensive technical knowledge. The
advantage of Panorama is in its efficient coding of information with the
principle of searching for information introduced in our solution being
very similar to human information processing within memory system.
Panorama's main goal is to facilitate the process of transferring
concepts from long-term memory to short-term memory by associations
\citep{ware_information_2004, atkinson_human_1968, waugh_primary_1965}
among concepts that user can freely traverse. The emphasis is given to
intuitive user interface \citep{obrenovic_adapting_2006} that enables
user with minimal technical skills to start browsing and building the
information space without extensive previous training. The ideas of
Memex
\citep{chakrabarti_using_2000, cole_visualizing_2002, bush_as_1945},
associations, context + focus techniques and particularly frames had a
profound influence on our notion of how Panorama was designed and
implemented. Panorama follows those concepts and presents an approach in
presenting data based on principles of associative thinking. In
employing Panorama approach for information system development several
good practices from existing object-oriented, rapid development and
peopleoriented methodologies were considered, while considering low
level of IT expertise required for development.

The remainder of this paper is structured as follows. In section
\ref{panorama-approach}, an overview of our approach to implementation
of information systems is given. The following section \ref{evaluation}
includes detailed comparison of Panorama approach and similar approaches
in different themes in information systems development. An experiment is
also presented that compares Panorama approach to object-oriented
approach on selected use cases. This is followed by the final section
that presents the conclusions and plans for future work.

\section{Panorama approach}\label{panorama-approach}

In the following section first architecture of Panorama will be
introduced with elements of data vocabulary and traversing the
information space. Then some aspects of Panorama, as input and reporting
module, will be discussed with other elements of panoramic view on data.
This section will also present information system development with
Panorama approach. After short comparison to existing approaches detail
steps will be depicted and their impact on the process.

The distinct value of Panorama approach is in supporting tool Panorama
that is used throughout the process -- in development and in using the
implemented system. When developing information system Panorama
facilitates definition of data vocabulary and enables automatic
transformation of logical model to implementation model and therefore
minimizes the required technical knowledge of users. By enabling this
transformation in Panorama approach users don't deal with programming
phase, but rather put more attention to analysis and design. After
successful implementation of information system, Panorama serves as
intuitive user interface with employing of design patterns
\citep{ahmed_model-based_2007} and enabling traversing the information
space by following associations and focus + context visualization
techniques that improve user experience.

\subsection{Data vocabulary and traversing the information
space}\label{data-vocabulary-and-traversing-the-information-space}

Panorama has a two-layer architecture. The top layer is the data
vocabulary containing all information about concepts being used. Data
vocabulary serves for creating underlying tables with all fields
corresponding to data vocabulary definition including fields that bind
tables with links. Data vocabulary characterizes objects of classes and
describes possible interconnections of objects from different classes.
Based on the content of data vocabulary Panorama ``interprets''
individual attributes and presents appropriate visualization to the
user.

Panorama's advantage is in \textbf{bidirectional nature of links} in
opposite to unidirectional links in documents for example found on the
internet. Panorama is using network data structure with all input and
output links indexed thus providing fast and elegant data retrieval.
User has the possibility to compare level of authority, i.e.~number of
incoming links, just by clicking neighbour records of the same class and
after choosing one of them he can enable filter on selected record to
broaden record which are shown in subsequent search.

\begin{figure}

{\centering \includegraphics[width=0.5\linewidth]{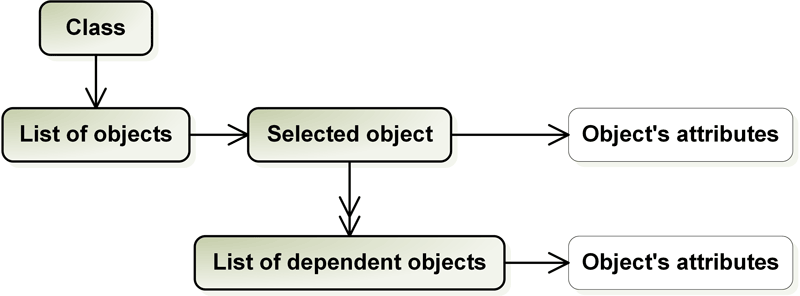} 

}

\caption{Panorama's 5 dimensional space}\label{fig:5D-space}
\end{figure}

\begin{figure}

{\centering \includegraphics[width=0.9\linewidth]{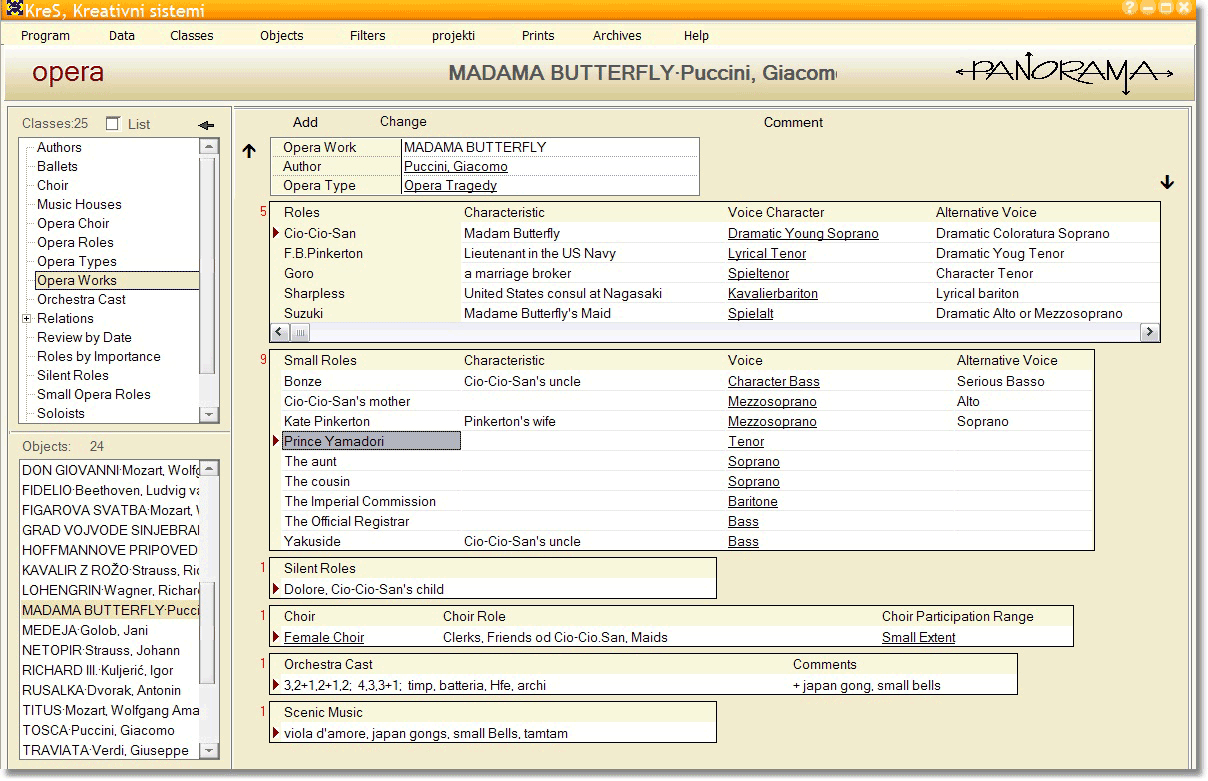} 

}

\caption{Panorama's user interface}\label{fig:Panorama-GUI}
\end{figure}

When using Panorama user deals with 5-dimensional space as depicted in
Figure \ref{fig:5D-space}. It follows a visual information-seeking
mantra for designers \citep{tergan_knowledge_2005}: overview first, zoom
and filter, then details-on-demand. User interface supports traversing
through this multi-dimensional space and is furthermore divided into two
parts (Figure \ref{fig:Panorama-GUI}):

\begin{itemize}
\tightlist
\item
  on the left hand side are selection windows for setting restrictions
  and
\item
  on the right hand side is the viewing window for displaying the
  observed object and its context.
\end{itemize}

\textbf{Selection windows} contain a list of \textbf{classes},
\textbf{objects} and \textbf{filters}. With the selection of a class
(\(1^{\text{st}}\) dimension), list of objects (\(2^{\text{nd}}\)
dimension) is simultaneously updated.

The class interconnection is the way of changing point of view of data.
Record attributes that belong to class interconnection data type are
implemented as links between data records. All links in Panorama are
bidirectional so the user can jump to records in both directions. Jump
to record pointed to by selected link opens new possibilities for link
selection and further data investigation. There is no need to go back in
the tree of selected choices to change the path of investigation as is
the case in tree data structure.

In Figure \ref{fig:Panorama-GUI} an example of opera works, developed by
one of the world famous conductors using Panorama tools is presented.
When one selects class Opera Works, the list of operas Don Giovanni,
Fidelio, Figaro's wedding and others is displayed. Furthermore, by
selecting the object (Madame Butterfly in our example), all the related
data within its context is displayed in \textbf{viewing window}. This
first of all includes object's attributes (\(3^{\text{rd}}\) dimension)
and the list of dependent objects (\(4^{\text{th}}\) dimension) with its
attributes (\(5^{\text{th}}\) dimension) respectively. Object's
attribute in our examples is Opera Work Madam Butterfly while dependant
objects are objects from various classes - Roles (Cio-Cio-San, F. B.
Pinkerton etc.), Small Roles (Bonze, Cio-Cio-San's mother etc.), Silent
Roles (Dolore, Cio-Cio-San's child), Choir (Female Choir), Orchestra
Cast and Scenic Music. All record links are displayed underlined.

Traversing from one information node to another is feasible simply by a
mouse click on a selected concept that consequently becomes the new
object of observation. As aforesaid one of the key elements of Panorama
are bidirectional links. They enable unlimited traversal across
information space. The observed object can also be marked as anchor
which denotes that, while traversing through related links is limited
only to selected object.

The starting point for all mechanisms in Panorama that facilitate
traversing from one object to another is in the design of specially
structured database that enables to establish n-way relationships
between objects and support object recognition process. Objects within
the context of observation are stored in the database and grouped into
classes of objects. All objects, instances of selected class, share the
same set of properties, but differ in values of attributes. Among
multi-value attributes, linking attributes can be found and they
represent a connection between object of observation and its
neighbourhood. Database used in Panorama is based on concepts of
relational databases (RDB)
\citep{codd_relational_1970, dietrich_advanced_2004} with
functionalities of object oriented databases (OODB)
\citep{kim_introduction_1990, bernstein_repositories_1998}. Access to
data in OODB can be faster because joins are often not needed (as in a
tabular implementation of a RDB). This is because an object can be
retrieved directly by following pointers and without a search. At the
ground level navigational databases approach
\citep{jutla_developing_1999} is followed which incorporates both the
network model and hierarchical model of database interfaces.
Navigational techniques use pointers and paths to navigate among data
records. This is in contrast to the relational model (implemented in
RDB), which strives to use declarative or logic programming techniques
where system is being queried for result instead of being navigated to.

Panorama enables a simple definition of relationships between objects,
without restrictions in quantity. Objects in Panorama can be linked in
the following ways:

\begin{itemize}
\item
  \textbf{1:1} or \textbf{1:n} relationship by a pointer attribute that
  is a member of source object's set of attributes. Panorama uses this
  type of relationship for defining \textbf{links}. For example:
  compositions can have the same title (Symphony No. 1), but different
  authors (Ludwig van Beethoven, Johannes Brahms etc.). When traversing
  through information space, Panorama offers a list of all known authors
  of selected compositions and on the other hand all compositions of
  each author are displayed regarding to the context.
\item
  \textbf{m:n} relationship can be established by an intermediate class
  or an object and in its simplest form the intermediate object contains
  a two pointer attribute, forming the concatenated key of the object.
\item
  \textbf{m:n:q} relationship (and subsequent cardinalities of
  relationships) can also be established by intermediate objects. The
  key is formed from concatenation of pointers and cardinality equals to
  number of relationships. Intermediate objects behave as normal objects
  and can be linked to subsequent relationships as basic objects can.
  They have their own identity and cannot be fully accessible through
  concatenated key. The multiple relationships with cardinality two or
  more represent the \textbf{information node}.
\end{itemize}

So far data vocabulary design of Panorama and structure of user
interface was presented, but this is only a prerequisite for computer
aided associative thinking process that is performed by the user.

\subsection{Panoramic view on data}\label{panoramic-view-on-data}

Besides controlled input of data and reporting module, Panorama also
offers user friendly, easy to use, intuitive interface for data review
and for traversing the information space. In ordinary applications all
these functionalities have yet to be programmed.

\begin{figure}

{\centering \includegraphics[width=0.5\linewidth]{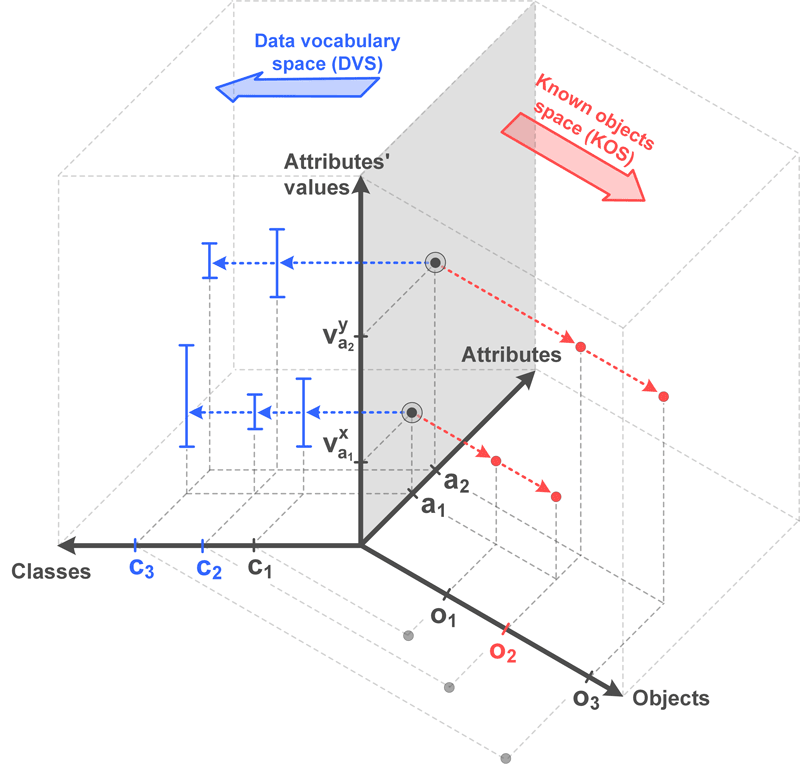} 

}

\caption{Object recognition process in information space}\label{fig:object-recognition}
\end{figure}

Computer aided associative thinking process supported in Panorama is
based on object recognition process (see Figure
\ref{fig:object-recognition}) that starts with observation and
perception of the environment. When observing the surrounding
environment the focus is on the specific object of observation,
nevertheless all the neighbouring objects are also comprehended.
Therefore the context of the observed object is identified by the set of
objects in the neighbourhood of that object. Classification of objects
into classes is based on the equality of attributes' sets (and interval
of values) that belong to respective objects from a set of objects.

The question that arises is what kind of mechanism to use for
neighbourhood observation and how to recognize each object based upon
their attributes? Obviously all information about classes, their
attributes and possible values for each of the attributes is stored
somewhere in our brains. This leads to the conclusion that we need a
very powerful data vocabulary, progressive by quantity of stored
information and by search methods and data classification. When a new
object is noticed, visible attributes of the object are perceived and we
try to find a match in our data vocabulary. The pattern of recognized
attributes defines the most probable class that the object belongs to.
In some cases further analysis of other attributes is needed to confirm
the correct class membership or to classify object into different class.

In Panorama object recognition is based on information residing in two
information spaces (see Figure \ref{fig:object-recognition}):

\begin{itemize}
\tightlist
\item
  \textbf{data vocabulary space (DVS)} and
\item
  \textbf{known object space (KOS)}.
\end{itemize}

The techniques used for recognition of new objects and traversing
through information space are based on several techniques in information
visualization \citep{ware_information_2004}, especially overview +
detail and focus + context approaches.

The simplicity of usage is one of the key elements of Panorama, because
target users are users with very little technical knowledge of details
in implementation. Querying complex data structures is based on
intuitive graphical interface, whereas Panorama does not use query
language to explore the information space, but it's rather based on
drag-and-drop approach by selecting objects with a mouse. This leads to
the following advantages:

\begin{itemize}
\tightlist
\item
  \textbf{User doesn't need to be familiar with the query language such
  as SQL} (similar approach to Query-by-Example and Visual Query Builder
  methods).
\item
  \textbf{There is no need for being familiar with the data structure},
  where in navigational approach users not need to know the names of
  appropriate attributes to build a query, because a user only follows
  associations.
\item
  For traversing the information space, \textbf{user} also \textbf{does
  not need to know the problem domain fully}.
\end{itemize}

A major disadvantage of using query languages is responding with ``no
result found'' message to user in case of a false condition. User is not
interested in what cannot be found, but only wants to know what is
available. With the use of Panorama and traversal that it supports it is
impossible to make a wrong turn to an empty information space.

The problems introduced are solved with intuitive user interface --
display of multiple concepts with filtered data that are in direct
relationship with observed object. Traversing through information does
not require entering commands, but only to follow the links that
represent a connection with linked objects. The traversal is feasible in
all directions and not limited by iteration. To facilitate quick start
usage of Panorama, several import options exist. Data can be imported
for selected class from multiple sources, while Panorama will recognize
the input structure and display it to the user who can take appropriate
actions. To maintain high level of adaptability, filters and anchors are
also available to help the user in coping with the large amount of
information and the appropriate formatting and displaying at latter to
the screen. Filter is used to simply restrict our traversal through
information space, so we can focus only on desired concepts. Anchor can
be set for multiple classes and it limits the displaying of objects of
selected classes only. For integration with other applications various
types (screen, printer, clipboard or file -- Excel, DBF, SQL, HTML, XML,
TXT, RTF, PDF) of exports are available. The common quality of creating
reports is the analogy with traversing the information space. Report can
simply be defined for selected object of observation and its context or
list of objects with applied restrictions. Again, no technical knowledge
is required in a form of entering query expressions, but simply by
selecting appropriate restrictions from data vocabulary.

\subsection{Panorama and IS
development}\label{Panorama-and-IS-development}

The general approach to developing information systems is the
Information Systems Development Life Cycle (SDLC)
\citep{rob_dilemma_2006, avison_information_2006}. Although there are
many variants, it has the following basic structure:

\begin{enumerate}
\def\labelenumi{\arabic{enumi}.}
\tightlist
\item
  feasibility study,
\item
  system investigation,
\item
  systems analysis,
\item
  systems design stage,
\item
  implementation and
\item
  review and maintenance.
\end{enumerate}

The use of methodology improves the practice of information systems
development, but due to dynamic environment and constantly changing
requirements it's hard to adopt only one approach with strictly defined
phases, techniques and tools. Panorama with its novel approach to
developing information systems follows examples mainly from
\textbf{Object-oriented methodologies}
\citep{rob_dilemma_2006, coad_object_1990, jacobson_unified_1999},
\textbf{Rapid development methodologies}
\citep{martin_rapid_1991, jeffries_extreme_2000, vidgen_developing_2002, milosavljevic_method_2004},
\textbf{People-oriented methodologies}
\citep{schreiber_knowledge_1999, griggs_web-based_2002, smaizys_business_2009}
and \textbf{Model-Driven approach}
\citep{gonzales-perez_modelling_2007, frankel_model_2003}.

Panorama also includes some aspects of Method Engineering (ME). ME,
sometimes referred to as methodology engineering, is the process of
designing, constructing, and merging methods and techniques to support
information systems development
\citep{brinkkemper_method_1996, vavpotic_approach_2009}. ME is
especially important in IS development as it is a sign of progress in
the adaptation of methodologies to changing IS/IT environments and
progress in methodology design itself. ME is often associated with the
hierarchical and bureaucratic approaches of the 1980s as techniques were
combined to form meta-methodologies. Normally it is concerned with the
blending of methods and techniques into a methodology or framework.
However, according to
\citep{avison_information_2006, henderson-sellers_situational_2010}, its
most recent form is Enterprise resource planning (ERP) systems, which
are combinations of application types rather than methods and
techniques. There are also other known advances in ME such as The OPEN
Process Framework (OPF) \citep{henderson-sellers_open_2000} and the
ISO/IEC 24744 international standard
\citep{iso/iec-24744_software_2007}.

The essential building blocks of information systems in object-oriented
methodologies are captured by defining objects -- both data and process
encapsulated together. The resemblance with using Panorama for
observation and perception of the environment can be found especially in
Object-Oriented Analysis (OOA). In OOA five major activities exists:
finding classes and objects, identifying structures, identifying
subjects, defining attributes and defining services. The end result of
finding classes and objects is a set of relevant classes, and for each
class the associated objects modelled using the appropriate conventions.
The second activity organizes basic classes and objects into hierarchies
that will enable the benefits of inheritance. Afterwards, identifying
subjects reduces the complexity of the model produced so far by dividing
or grouping into more manageable and understandable subject areas.
According to OOA, objects are composed of data and processing, where
defining attributes defines the data and defining services defines the
processing. The similar process is also conducted when building data
vocabulary in Panorama. Implementation (step 5) and review and
maintenance (step 6) in SDLC represent the majority of the time spent on
project development \citep{lajovic_panoramic_2005}. To successfully
complete projects that don't require a lot of calculation and don't
emphasize process oriented aspects (e.g.~office management, CRM or
record management), Panorama offers a fast and efficient way. The first
four phases (feasibility study, system integration, systems analysis and
systems design) from SDLC are always present, because the development is
not feasible without presence of requirements and designs. Panorama on
the other hand tries to minimize the impact of phases 5 and 6 on the
time required for completing the project. Our solution contains required
mechanisms for updating and printing the content of data vocabulary and
also controlled input of data with referential integrity checking. This
leads to avoidance of the major part of the work that has to be done in
implementation and review and maintenance phase. Furthermore the vast
majority of complications are also avoided:

\begin{itemize}
\tightlist
\item
  The lack of communication between designers and programmers leads to
  misunderstandings.
\item
  Data model doesn't cover all the requirements.
\item
  The final application is not fully tested and it contains several
  bugs.
\item
  Subsequent changes to the application, based on the requirements
  defined after the completion of the project etc.
\end{itemize}

Guidelines similar to those of Panorama's can also be found in Rapid
development methodologies, whereas Rapid Application Development (RAD)
is a response to the need to develop information systems more quickly
\citep{avison_information_2006} because of rapidly changing business
requirements. James Martin's RAD (JMRAD) and Web IS development
methodology (WISDM) place emphasis on the early stages of systems
development as is the case of using Panorama. This concerns the
definition of requirements, analysis and design with the use of a
language that is of the business and the users, rather than the more
technical language of information systems. As mentioned before, common
guidelines can also be found in People-oriented methodologies,
particularly in KADS, CommonKADS and End-user computing (EUC).
People-oriented methodologies encompass the socio-technical view that,
in order to be effective, the technology must fit closely with the
social and organizational factors in the application domain. These
methodologies are people oriented in the sense that they attempt to
capture the expertise and knowledge of people in the organization. They
emphasize the importance of knowledge management and leverage knowledge
as an important organizational resource. Knowledge, the ability to use
information in action for a particular purpose, is even more important
than information, which is more commonplace. Panorama in some aspects
follows End-user computing paradigm, found in people oriented themes,
where application is developed by non-specialist IT people, particularly
in the form of spread-sheets. Main advantages are an effective way of
providing some functionality of the information systems in organizations
and an increase in user satisfaction. As Method Engineering emphasizes
usage of best practices from systems development, Panorama can also be
analysed using this perspective, but it is limited to data oriented
approaches. While Method Engineering approaches (i.e.~OPF) mainly focus
on the process (framework, libraries, reusable methods, usage
guidelines), thou also mention tools, Panorama is rather focused on the
tool support that follows proposed approach.

We can therefore conclude that Panorama follows a kind of
knowledgeengineering methodology with iterative approach and following
next steps:

\begin{enumerate}
\def\labelenumi{\arabic{enumi}.}
\tightlist
\item
  Determination of the domain,
\item
  Identification and grouping of information (project definition),

  \begin{enumerate}
  \def\labelenumii{\arabic{enumii}.}
  \tightlist
  \item
    Assembling of important concepts,
  \item
    Defining classes,
  \item
    Defining attributes and relationships and
  \end{enumerate}
\item
  Using the project.
\end{enumerate}

We start with defining the domain and scope (step 1) which coincides
with step 1 from SDLC. This includes the determination of the problem
domain, type of questions that the Panorama model should provide answers
to, users responsible for using and maintaining data vocabulary and
known objects space etc.

Step 2 deals with construction of data vocabulary that characterizes
objects of classes and describes possible interconnections of objects
from different classes. Based on the content of data vocabulary,
Panorama interprets individual attributes and presents appropriate
visualization to the user. Defining data vocabulary is furthermore
consisted of three sub steps: assembling of important concepts, defining
the classes and defining the properties.

In assembling of important concepts the main concepts are identified,
what properties those concepts have and what would we like to say about
them. It is important to get a comprehensive list of concepts without
paying special attention to overlaps between concepts they represent,
relationships among the concepts and properties.

The next action is to define the classes which are used to group objects
with the same properties. When classes are described they alone don't
provide enough information to answer the questions from step 1,
therefore we must describe the internal structure of concepts.

The phase of defining classes has already selected classes from the list
created in step 2.1. Most of the remaining concepts are likely to be
attributes of these classes, mandatory or optional. Besides these
proprietary attributes, each object also consists of a set of additional
attributes that position selected object in space and time. By defining
relationships among concepts we define foundation for Panorama to enable
traversing through information space. Actions that have been made so far
have roughly covered steps 2 to 5 in SDLC and prepared all the necessary
requirements to start using Panorama. By using Panorama we consider
entering new data or defining known objects space (step 3). This
requires selecting a class, creating individual instance of that class
and filling in perceived attributes.

Panorama tool offers users complete support in using this approach. In
initial steps users define data vocabulary with all the restrictions
that apply. The tool supports building data vocabulary from scratch and
also importing data from several resources. When data vocabulary is
defined user can enter all the restrictions that apply and finally
couple this schema with actual data again from scratch or from various
data sources. At this stage the use of the project starts where Panorama
deals with all translations from abstract concepts in data vocabulary to
actual implementation at the information system level. The traversal
through information space is supported by intuitive user interface which
does not require users to learn query language.

\section{Evaluation}\label{evaluation}

In the evaluation part of this research we tackled the following
research questions:

\begin{enumerate}
\def\labelenumi{\arabic{enumi}.}
\tightlist
\item
  How does developing computer software with Panorama approach differs
  from existing, especially object-oriented approaches?
\item
  What are the problem domains where application of Panorama approach is
  the most feasible?
\item
  What is the impact of Panorama approach regarding time and technical
  knowledge of users for completing the software development life cycle?
\end{enumerate}

First a comparison between Panorama approach, STRADIS, SSADM, IE, DSDM,
OOA and RUP was conducted following framework for comparing
methodologies defined in \citep{avison_information_2006}. Based on the
findings from that comparison an experiment was carried out with \(4\)
experts and \(24\) students as raters, who rated \(2\) problem domains
and \(2\) approaches. Raters heuristically evaluated the required a
priori knowledge for each of the approaches and time required to
complete individual phase was measured.

\subsection{Comparison of approaches}\label{comparison-of-approaches}

\subsubsection{Method}\label{method}

We are aware that Panorama approach is not a methodology, because a
methodology encompasses much more than an approach. Nevertheless the
framework mentioned before is still appropriate due to its very generic
design. There are several questions that need to be answered to give a
complete comparison:

\begin{itemize}
\tightlist
\item
  What are the underlying philosophical assumptions of the methodology?
  What makes it a legitimate approach?
\item
  What particular skills are required of the participants?
\item
  What representation, abstractions, and models are employed?
\item
  What aspects of the development process does the methodology cover?
\item
  What is the focus of the methodology? Is it, for example, people-,
  data-, process-, and/or problem-oriented? Does it address
  organizational and strategic issues?
\end{itemize}

Besides Panorama approach six other approaches from different themes in
information systems development were considered for comparison as
follows:

\begin{itemize}
\tightlist
\item
  \textbf{Structured Analysis, Design and Implementation of Information
  Systems (STRADIS)} from process-oriented methodologies,
\item
  \textbf{Structured Systems Analysis and Design Method (SSADM)} from
  blended methodologies,
\item
  \textbf{Information Engineering (IE)} from blended methodologies,
\item
  \textbf{Dynamic Systems Development Method (DSDM)} from rapid
  development methodologies,
\item
  \textbf{Object-Oriented Analysis (OOA)} from object-oriented
  methodologies and
\item
  \textbf{Rational Unified Process (RUP)} from object-oriented
  methodologies.
\end{itemize}

Each of the seven selected approaches was discussed and mutually
compared according to elements of the framework for comparing
methodologies.

\subsubsection{Results}\label{results}

There are number of sub elements to \textbf{philosophy} which we will
examine in turn. The first sub element is paradigm where science and
systems paradigm are identified as to be of critical importance. In
science paradigm complexity is handled through reductionism, breaking
things down into smaller and smaller parts for examination and
explanation. In systems paradigm concern is for the whole picture, the
emergent properties (e.g.~the whole is greater than the sum of parts),
and the interrelationships between parts of the whole. All approaches
(STRADIS, SSADM, IE, DSDM, OOA, RUP and Panorama) follow science
paradigm with clear reductionist approach and they all accept the
ontological position of realism. They all argue that universe comprises
objectively given, immutable objects and structure, whereas these exist
as empirical entities, on their own, independent of the observer's
appreciation of them. One of fairly obvious clue to the methodology
philosophy is the stated objectives. DSDM, although often resulting in
the design of computer systems, is sometimes used to address
organizational or general problemsolving issues. STRADIS, SSADM, IE, OOA
and RUP all claim that they are not general problem-solving
methodologies, but have clear objectives to develop computerized
information systems. Panorama is also not general problem-solving
approach, but its objective is to combine very heterogeneous information
into a stable and networked computerized information system. Very
important aspect is to enable display of object and its context and
enable two-way interactions with objects from the context. The third
factor relating to philosophy is the domain of situations that
methodologies address. IE is identified as being of the planning,
organization, and strategy type. It is an approach adopting the
philosophy that an organization needs a strategic plan in order to
function effectively, and that success is related to the identification
of information systems that will benefit the organization and help
achieve its strategic objectives. STRADIS, SSADM, DSDM, OOA and RUP are
classified as specific problem-solving methodologies; i.e.~they do not
focus on identifying systems required by the organization but begin by
assuming that a specific problem is to be addressed. Panorama can like
others also be classified as specific problem-solving methodologies with
distinct top-down approach and identification of organization's focus.
It is also important to notice that Panorama is not intended for process
oriented domains. Most methodologies appear to claim to be general
purpose. OOA and RUP are considered to be general purpose, although is
suggested that they are not very helpful for simple, limited systems or
systems with only a few class \& objects. STRADIS is also stated to be
general purpose and applicable to any size of system, yet the main
technique is data flow diagramming, which is not particularly suitable
for all types of application, for example, the development of management
information systems or web-based systems. SSADM and also STRADIS, IE,
OOA and RUP have all been designed primarily for use in large
organizations. Panorama is intended to help develop information systems
in smaller environments or where the target system is PC-based. It is
particularly suitable for the following types of application -- office
management, CRM, record management, planning, maintenance; and various
user types, including management level.

The second element of the framework, the \textbf{model}, can be
investigated in terms of the type of model, the levels of abstraction of
the model, and the orientation or focus of the model. The primary
process model used is the data flow diagram and it can be found in
STRADIS, while in SSADM is an important model, although not the only
one. It is also present in IE, but play a less significant role than,
for example, in STRADIS. The data flow diagram is predominantly a
process model, and data are only modeled as a by-product of the
processes. In OOA and RUP the basic models are in integration of both
process and data orientation, often in the same diagram, which is a key
element of the object-oriented approach. Panorama in contrast to other
approaches uses data-oriented modeling approach such as ER model whereas
process orientation is not handled as an important aspect.

The third element of the framework is that of the \textbf{techniques and
tools} that a methodology employs. STRADIS is an example of a
methodology which is largely described in terms of its techniques. A
methodology which advocates the clear separation of the modeling of data
and processes is SSADM. OOA and RUP use object-oriented techniques and
models or models that incorporate the essential combination of data and
process. Tools range from simple drawing tools through to tools
supporting the whole development process, including prototyping, project
management, code generation, simulation and so on. IE explicitly suggest
that the techniques are not fundamental part of the methodology and that
the current recommended techniques can be replaced. Panorama approach is
mainly presented in a view of a tool that is used in throughout the
development process by developers and users and as a final product by
users. It supports the complete lifecycle of software development and is
also deployed to a client and plays a vital role as a toolset which
encompasses model, editor, reports, graphics screen interfaces, code and
database generators, configuration tools etc.

\textbf{Scope} is an indication of the stages of the life cycle of stems
development which the methodology covers (\(0\) -- it does not mention,
\(1\) -- briefly mentioned, \(2\) -- address the area, \(3\) -- covers
the stage in some detail) and is presented in the following Table
\ref{tab:scope-comparison}.

\begin{table}

\caption{\label{tab:scope-comparison}Scope comparison of different software development approaches}
\centering
\begin{tabular}[t]{lccccccc}
\toprule
Stage & STRADIS & SSADM & IE & DSDM & OOA & RUP & Panorama\\
\midrule
Strategy & 0 & 2 & 3 & 2 & 0 & 1 & 0\\
Feasibility & 3 & 3 & 2 & 3 & 0 & 2 & 1\\
Analysis & 3 & 3 & 3 & 3 & 3 & 3 & 3\\
Logical design & 3 & 3 & 3 & 3 & 3 & 3 & 3\\
Physical design & 3 & 3 & 3 & 3 & 2 & 3 & 3\\
\addlinespace
Programming & 2 & 0 & 2 & 2 & 0 & 2 & 3\\
Testing & 2 & 0 & 3 & 2 & 0 & 3 & 2\\
Implementation & 1 & 2 & 2 & 2 & 0 & 3 & 3\\
Evaluation & 1 & 1 & 2 & 0 & 0 & 2 & 0\\
Maintenance & 0 & 0 & 1 & 0 & 0 & 0 & 1\\
\bottomrule
\end{tabular}
\end{table}

The main focus of most compared methodologies is at the analysis and
design stages, while maintenance is by far the worst covered stage.
Panorama is also very strong in analysis and design phase but it also
addresses maintenance.

The next element in the framework evaluates the \textbf{deliverables} at
each stage and, in particular, the nature of the final deliverable. This
can vary from being an analysis specification to a working
implementation of a system and all its related procedures. STRADIS,
SSADM, IE, DSDM, OOA and RUP all include analysis and design
specification, while STRADIS, SSADM, IE, DSDM and RUP also cover working
implementation of a system. In Panorama approach the emphasis is on
working implementation of a system while products at the end of phases
include a working version of logical model and data vocabulary.

The background of the methodology broadly identifies its origins in
terms of academic or commercial. STRADIS, SSADM, IE, DSSDM, OOA and
Panorama lie in the commercial sphere, whereas RUP has academic
backgrounds. There is a view that commercial methodologies are not in as
widespread use as is claimed but finding evidence is difficult.
According to a research \citep{avison_information_2006} \(57\%\) of
organizations were using a systems development methodology, but, of
these, only \(11\%\) used a commercial development methodology, \(30\%\)
used a commercial methodology adapted for in-house use, and \(59\%\) a
methodology which was internally developed and not based on an a
commercial methodology. Panorama approach has a wide user base among
different levels in organizations. Because of its design it is
interesting to point out that is especially suitable for less
technically oriented users.

The traditional view of information systems development is that a
specialist team of professional systems analysts and designers perform
the analysis and design aspects and professional programmers design the
programs and write the code. Although the exact roles have different
names, in general systems development is undertaken by professional
technical developers. This view is taken by STRADIS, SSADM, IE, DSDM,
OOA and RUP. Panorama takes a different view, and users have a much more
active role by following JAD (Joint Application Design) which includes
domain experts and professional technical developers. Therefore in
Panorama, the users are directly involved in the analysis and design
with the professional analysts as consultants. It has turned out, as
seen in the following experiment, that leaving out analysis of current
state could have beneficial consequences. In almost all methodologies
considerable training and experience is necessary for at least some of
the players. This may significantly increase the time and costs required
to develop a project. With Panorama time required for implementation is
very short because majority of features are automatically built from
data vocabulary.

The last element of the framework is what purchasers actually get for
their money. This may consist of software, written documentation, an
agreed number of hours' training, a telephone help service, and
consultancy etc. With SSADM the product is large and copious sets of
manuals. RUP has a range of documents, books, and specifications but
also has a multimedia website, and indeed claims that the methodology
product is actually delivered using this web technology. Panorama
approach produces two types of products:

\begin{itemize}
\tightlist
\item
  a tool that facilitates software development and is at later stage
  also used on a client side as implementation and
\item
  the project, based on JAD approach together with documentation.
\end{itemize}

\subsubsection{Discussion}\label{discussion}

The following table summarizes the view about comparing selected
methodologies. It can be argued that several differences exist among
approaches, which is expected, because of their origin in different
themes in information systems development.

Panorama stands out especially because of its data-oriented model and
very strong orientation into a working implementation of a system with
object and its context visualization approaches (see Table
\ref{tab:SDA-comparison}).

\begin{table}

\caption{\label{tab:SDA-comparison}Comparison of software development approaches (abbreviations used are further explained in the text, following the table)}
\centering
\begin{tabular}[t]{rlccccccc}
\toprule
  &   & STRADIS & SSADM & IE & DSDM & OOA & RUP & Panorama\\
\midrule
\textbf{1.} & \textbf{Philosophy} &  &  &  &  &  &  & \\
a) & Paradigm & Pa-1 & Pa-1 & Pa-1 & Pa-1 & Pa-1 & Pa-1 & Pa-1\\
b) & Objectives & Ob-1 & Ob-1 & Ob-1 & Ob-2 & Ob-1 & Ob-1 & Ob-3\\
c) & Domain & Do-1 & Do-1 & Do-2 & Do-1 & Do-1 & Do-1 & Do-3\\
d) & Target & Ta-1 & Ta-1 & Ta-1 & Ta-1 & Ta-1 & Ta-1 & Ta-2\\
\addlinespace
\textbf{2.} & \textbf{Model} & Mo-1 & Mo-2 & Mo-2 & Mo-3 & Mo-4 & Mo-4 & Mo-5\\
\textbf{3.} & \textbf{Techniques and tools} & TT-1 & TT-2 & TT-3 & TT-4 & TT-5 & TT-6 & TT-7\\
\textbf{4.} & \textbf{Scope} & Sc-1 & Sc-2 & Sc-3 & Sc-4 & Sc-5 & Sc-6 & Sc-7\\
\textbf{5.} & \textbf{Outputs} & Ou-1 & Ou-1 & Ou-1 & Ou-1 & Ou-1 & Ou-1 & Ou-2\\
\textbf{6.} & \textbf{Practice} &  &  &  &  &  &  & \\
\addlinespace
a) & Background & Bg-1 & Bg-1 & Bg-1 & Bg-1 & Bg-1 & Bg-2 & Bg-1, Bg-2\\
b) & User base & Us-1 & Us-1 & Us-1 & Us-1 & Us-1 & Us-1 & Us-2\\
c) & Participants & Ba-1 & Ba-1 & Ba-1 & Ba-1, Ba-2 & Ba-1 & Ba-1 & Ba-1, Ba-2\\
\textbf{7.} & \textbf{Product} & Pr-1 & Pr-2 & Pr-1 & Pr-1 & Pr-1 & Pr-3 & Pr-4\\
\bottomrule
\end{tabular}
\end{table}

Abbreviations used in the Table \ref{tab:SDA-comparison} are as follows:

\begin{itemize}
\tightlist
\item
  \textbf{Paradigm}:

  \begin{itemize}
  \tightlist
  \item
    \texttt{(Pa-1)} Science paradigm, clear reductionist approach,
    acceptance of the ontological position of realism.
  \end{itemize}
\item
  \textbf{Objectives}:

  \begin{itemize}
  \tightlist
  \item
    \texttt{(Ob-1)} Not a general problem-solving methodology, but as
    having clear objectives to develop computerized information systems.
  \item
    \texttt{(Ob-2)} Often resulting in the design of computer systems,
    but is still sometimes used to address general problem-solving
    issues.
  \item
    \texttt{(Ob-3)} Not a general problem-solving approach, but to
    combine very heterogeneous information into networked IS.
  \end{itemize}
\item
  \textbf{Domain}:

  \begin{itemize}
  \tightlist
  \item
    \texttt{(Do-1)} Specific problem-solving methodologies. Do not focus
    on identifying the systems required by the organization but begin by
    assuming that a specific problem is to be addressed.
  \item
    \texttt{(Do-2)} Planning, organization and strategy type, where
    organization needs a strategic plan in order to function
    effectively.
  \item
    \texttt{(Do-3)} Specific problem-solving approach. Not intended for
    process oriented domains.
  \end{itemize}
\item
  \textbf{Target}:

  \begin{itemize}
  \tightlist
  \item
    \texttt{(Ta-1)} General purpose, primarily designed for use in large
    organizations,
  \item
    \texttt{(Ta-2)} Smaller environments, especially office management,
    CRM, record management, planning, project management, maintenance
    etc.
  \end{itemize}
\item
  \textbf{Model}:

  \begin{itemize}
  \tightlist
  \item
    \texttt{(Mo-1)} Data flow diagram is the primary process model,
  \item
    \texttt{(Mo-2)} Data flow diagram is an important (although not only
    one) model,
  \item
    \texttt{(Mo-3)} Usually data flow diagram and data oriented model,
  \item
    \texttt{(Mo-4)} The basic model is an integration of both process
    and data orientation, often in the same diagram, which is a key
    element of the object-orient approach.
  \item
    \texttt{(Mo-5)} Data-oriented model.
  \end{itemize}
\item
  \textbf{Techniques and tools}:

  \begin{itemize}
  \tightlist
  \item
    \texttt{(TT-1)} Largely described in terms of techniques,
  \item
    \texttt{(TT-2)} Clear separation of modeling of data and processes.
    Recommend the tools to some degree.
  \item
    \texttt{(TT-3)} Techniques are not fundamental part.
  \item
    \texttt{(TT-4)} No specific tools are recommended,
  \item
    \texttt{(TT-5)} OO techniques. Tools might be helpful but are not
    necessarily essential.
  \item
    \texttt{(TT-6)} OO techniques. Process should not be contemplated
    without the use of tools, the process being too complicated and time
    consuming.
  \item
    \texttt{(TT-7)} ER model as primary technique and mainly presented
    in a view of a tool Panorama.
  \end{itemize}
\item
  \textbf{Scope}:

  \begin{itemize}
  \tightlist
  \item
    \texttt{(Sc-1)} Feasibility, analysis, logical and physical design
    are covered in detail. Programming and testing are addressed.
    Implementation and evaluation are only briefly mentioned.
  \item
    \texttt{(Sc-2)} Feasibility, analysis, logical and physical design
    are covered in detail. Strategy and implementation are addressed.
    Evaluation is only briefly mentioned.
  \item
    \texttt{(Sc-3)} Strategy, analysis, logical and physical design and
    testing are covered in detail. Feasibility, programming,
    implementation and evaluation are addressed. Maintenance is briefly
    mentioned.
  \item
    \texttt{(Sc-4)} Feasibility, analysis, logical and physical design
    are covered in detail. Strategy, programming, testing and
    implementation are addressed.
  \item
    \texttt{(Sc-5)} Analysis and logical design are covered in detail.
    Physical design is addressed.
  \item
    \texttt{(Sc-6)} Analysis, logical and physical design, testing,
    implementation and evaluation are covered in detail. Feasibility and
    programming are addressed. Strategy is only briefly mentioned.
  \item
    \texttt{(Sc-7)} Analysis, logical and physical design, programming,
    and implementation are covered in detail. Testing is addressed.
    Feasibility and maintenance are only briefly mentioned.
  \end{itemize}
\item
  \textbf{Outputs}:

  \begin{itemize}
  \tightlist
  \item
    \texttt{(Ou-1)} Analysis and design specification and working
    implementation of a system and
  \item
    \texttt{(Ou-2)} Working version of logical model and data vocabulary
    and working implementation of a system.
  \end{itemize}
\item
  \textbf{Background}:

  \begin{itemize}
  \tightlist
  \item
    \texttt{(Bg-1)} Commercial and
  \item
    \texttt{(Bg-2)} Academic.
  \end{itemize}
\item
  \textbf{User base}:

  \begin{itemize}
  \tightlist
  \item
    \texttt{(Us-1)} Various and numerous and
  \item
    \texttt{(Us-2)} Small.
  \end{itemize}
\item
  \textbf{Participants}:

  \begin{itemize}
  \tightlist
  \item
    \texttt{(Ba-1)} Professional technical developers and
  \item
    \texttt{(Ba-2)} Business users.
  \end{itemize}
\item
  \textbf{Product}:

  \begin{itemize}
  \tightlist
  \item
    \texttt{(Pr-1)} Documentation,
  \item
    \texttt{(Pr-2)} Large and copious sets of manuals,
  \item
    \texttt{(Pr-3)} Range of documents, books, specifications and a
    multimedia website and
  \item
    \texttt{(Pr-4)} Tool, project implementation and documentation.
  \end{itemize}
\end{itemize}

Based on these findings an experiment was conducted to measure the
effectiveness of object-oriented (OO) and Panorama approach on two
selected domains, as those approaches share the most commonalities and
OO approach is one of the most widespread in information system
development.

\subsection{Experiment}\label{experiment}

\subsubsection{Method}\label{method-1}

Altogether \(28\) members were chosen to conduct the experiment. \(4\)
members (\(\boldsymbol{P_1}\) to \(\boldsymbol{P_4}\)) from University
of Ljubljana, Faculty of Computer and Information Systems, Information
Systems laboratory and \(24\) students (\(\boldsymbol{S_1}\) to
\(\boldsymbol{S_{24}}\)) that attended undergraduate program at the same
University. Evaluators \(P_1\) to \(P_4\) were experts in software
development and had extensive practical experience with object-oriented
software development, while evaluators \(S_1\) to \(S_{24}\) were users
without extensive technical knowledge and inexperienced in software
development. Panorama approach was presented to all evaluators before
conducting the experiment, because none of them had any previous
experience.

Before engaging the experiment all evaluators were randomly divided into
two groups \(A\) and \(B\) of equal size and the same ratio among
experts and students. This resulted in individual groups containing
\(2\) expert members and \(12\) students. Each of the group had to
evaluate both Object-oriented and Panorama approach, but in turn in
different problem domain.

For evaluation purposes two fairly simple use cases \(C_1\) and \(C_2\)
were identified. Use case \(\boldsymbol{C_1}\) is concerned about
\textbf{information system support for cinemas}, whereas information
about cinema halls, timetables, movies, genres and customer's preferred
seats by each cinema hall has to be available. Several functionalities
had to be implemented, including performing analyses by customer to see
which movie genre are they interested in, when they go to the cinema
etc. Use \(C_2\) deals with \textbf{information support for mobile
operators billing}. Support for bundled services, customers, packages,
bill detail types, conversation rates and subscription fees had to be
captured. Master-detail view on the expenses related to customer has to
be supported to enable further analyses of rating information.

Each group of participants (\(A\) and \(B\)) was requested to implement
use cases \(C_1\) and \(C_2\) using Object-oriented and Panorama
approach. The requirements of cases \(C_1\) and \(C_2\) were described
in detail with the natural language. Participants had a goal to produce
a working computer program following given use case and accompanying
documentation. In the last phase of Software Development lifecycle --
Evaluation, all results from participants were independently
double-blind evaluated by domain experts according to the original
objectives and requirements. If these requests were not met, users were
required to repair their solution and if afterwards the solution was
still inadequate, the participant (\(P_1\) to \(P_4\) and \(S_1\) to
\(S_{24}\)) was excluded from the analysis. Due to this issue three
participants were excluded from the set of raters which resulted in
reduced number of participants from \(27\) to \(24\).

With Object-oriented approach PowerDesigner tool was used in analysis
and design phase for user requirements analysis and logical design of
the system. UML modeling notation was used to capture and define the
problem domain (use case diagrams, class diagrams etc.). Eclipse
development environment with a Microsoft SQL Server database was used in
the programming phase. For Panorama approach Panorama tool was used
throughout the software development cycle, following steps defined in
section \ref{Panorama-and-IS-development}.

The experiment was then performed in two steps (see Table
\ref{tab:experiment-procedure}). First group \(A\) implemented case
study \(C_1\) using Object-oriented approach, while group \(B\)
implemented case study \(C_1\) using Panorama approach. In second step
roles have been reversed and group \(B\) implemented \(C_2\) using
object-oriented approach and group \(A\) implemented \(C_2\) using
Panorama approach.

\begin{table}

\caption{\label{tab:experiment-procedure}Procedure of conducting experiment}
\centering
\begin{tabular}[t]{lcc}
\toprule
  & Object-oriented approach & Panorama approach\\
\midrule
Step 1: $\textbf{Use case }\boldsymbol{C_1}$ & group $\boldsymbol{A}$ & group $\boldsymbol{B}$\\
Step 2: $\textbf{Use case }\boldsymbol{C_2}$ & group $\boldsymbol{B}$ & group $\boldsymbol{A}$\\
\bottomrule
\end{tabular}
\end{table}

For each participant several elements were measured, from time required
to complete each stage of software development lifecycle, final outputs
and artifacts between the stages and at the end whether the requirements
were met.

The phases of the Software Development approach were evaluated on a time
consumption and output basis. In analysis users were required to get
familiar with problem domain and no special outputs were requested from
this phase. In logical design users had to capture the requirements in
selected notation (UML use case and class diagrams for Object-oriented
approach and business vocabulary in Panorama). In physical design
transformation of logical into physical model is required when specific
environment is selected. The outputs required from this phase include
generation of database schema in Object-oriented approach. In
programming phase users were requested to implement required
functionalities and the output is a working prototype. The following
phase of testing deals with implemented prototype and leads to
elimination of any development errors. As already mentioned, in
evaluation results are compared to original requirements.

Besides measurements, the questionnaire (Table
\ref{tab:questionnaire-outline}) was introduced to give feedback on
different approaches used for information system development. The aim
was to capture users view on important aspects such as capability of
approach to express static and dynamic aspects of the system, the
required effort to progress from design to implementation stage, level
of user participation etc. All questions were in Liker-type format.

\begin{table}

\caption{\label{tab:questionnaire-outline}Questionnaire for measuring the development process}
\centering
\begin{tabu} to \linewidth {>{\raggedright}X>{\centering\arraybackslash}p{3cm}>{\centering\arraybackslash}p{2cm}}
\toprule
  & Time spent [min] & Outputs\\
\midrule
\textbf{Analysis} &  & \\
Includes user requirements analysis. &  & \\
\textbf{Logial design} &  & \\
Design independent of physical environment. &  & \\
\textbf{Physical design} &  & \\
\addlinespace
\textbf{Programming} &  & \\
Physical development of the system. &  & \\
\textbf{Testing} &  & \\
Planning as well as the testing of systems, programs, and procedures. &  & \\
\textbf{Implementation} &  & \\
\addlinespace
Planning and implementation of technical, social, and organizational aspects. &  & \\
\textbf{Evaluation} &  & \\
Measurement and evaluation of the implemented system and a comparison with the original objectives. &  & \\
\bottomrule
\end{tabu}
\end{table}

The aim of the experiment was to investigate following hypotheses
presented in Table \ref{tab:hypotheses} and after collecting the data
the analysis was performed using SPSS toolkit.

\begin{table}

\caption{\label{tab:hypotheses}Hypotheses of the experiment}
\centering
\begin{tabu} to \linewidth {>{\raggedleft\arraybackslash}p{1cm}>{\raggedright}X}
\toprule
$\boldsymbol{H_1}$ & Time spent to complete software development of selected use cases is shorter with Panorama than object-oriented approach.\\
$\boldsymbol{H_2}$ & Panorama puts emphasis on analysis, logical design and testing, while objectoriented approach on physical design and programming.\\
$\boldsymbol{H_3}$ & Panorama is more appropriate for domains with static (data oriented) rather than dynamic (process oriented) components.\\
$\boldsymbol{H_4}$ & With Panorama approach in contrast to object-oriented approach, users are more encouraged to participate, required technical knowledge for development is lower, introducing additional functionalities is less demanding and the approach is easier to learn.\\
\bottomrule
\end{tabu}
\end{table}

\subsubsection{Results}\label{results-1}

A total of \(2\) case studies (\(C_1\) and \(C_2\)), \(28\) raters
(\(P_1\) to \(P_4\) and \(S_1\) to \(S_{24}\)) and \(2\) approaches
(Panorama and Object-oriented) were included in the experiment.
Aforementioned resources resulted in \(56\) executions using different
approaches on different use cases \((2 \times 2 \times 14)\).

To test for differences in rater bias ANOVA (ANalysis Of VAriance) model
was used, including the calculation of intraclass correlation (ICC) and
to describe raters' marginal distributions graphical method of
histograms was employed.

With intraclass correlation (ICC)
\citep{shrout_intraclass_1979, yaffee_enhancement_1998, hart_evaluating_2008}
rating reliability was assessed by comparing the variability of
different ratings of the same subject (questions from the questionnaire
and effort distributions on Software Development activities by approach)
to the total variation across all ratings and all subjects. ICC is
calculated according to the following formula

\begin{equation}
ICC = \frac{\sigma^2(subjects)}{\sigma^2(subjects) + \sigma^2(raters)}
\label{eq:ICC}
\end{equation}

Because in practice we do not know the true values of
\(\sigma^2(subjects)\) and \(\sigma^2(raters)\), we must instead
estimate them from sample data. With calculation of ICC we have to be
aware that there exist several types - Case 1, Case 2 and Case 3. For
our experiment Case 2 was selected, where the same set of \(k\) raters
rate each subject. This corresponds to a fully-crossed
\((rater \times subject)\), \(2\)-way ANOVA design in which both
\(subject\) and \(rater\) are separate effects. Because rater is
considered a random effect, this means that \(k\) raters in the study
are considered a random sample from a population of potential raters;
therefore this type \(2\) of ICC estimates the reliability of the larger
population of raters.

\paragraph{\texorpdfstring{Hypothesis
\(\text{H}_1\)}{Hypothesis \textbackslash{}text\{H\}\_1}}\label{hypothesis-texth_1}

The claim of \(H_1\) is that time spent to complete software development
of selected use cases is shorter with Panorama than Object-oriented
approach. The hypothesis was tested with \(2\)-way ANOVA model with
factors \emph{Approach} and \emph{Case study} and dependent variable
\emph{Total time}. Before conducting ANOVA test, exploratory data
analysis was performed to confirm that data is normally distributed, so
first pre-requirement to proceed with ANOVA was met. Levene's test of
equality of error variances also indicated that the error variance of
the dependent variable is equal across the groups, i.e.~the assumption
of the ANOVA test has been met.

The results of \(2\)-way ANOVA test are depicted in Table
\ref{tab:h1-anova}, where is clearly seen that the only significant
effect on Total time variable is of the factor \emph{Approach}, while
factor \emph{Case study} and \emph{Approach together with Case study}
don't have a significant effect on \emph{Total time}.

\begin{table}

\caption{\label{tab:h1-anova}Tests of between-subjects effects on dependent variable Total time}
\centering
\begin{tabular}[t]{lrrc}
\toprule
  &   & F-value & Sig.\\
\midrule
$\text{Corrected model}$ & $F(3,55)=$ & $46,44$ & $\checkmark$\\
$\text{Intercept}$ & $F(1,55)=$ & $6.357,42$ & $\checkmark$\\
$\textbf{Approach}$ & $F(1,55)=$ & $\boldsymbol{137,48}$ & $\checkmark$\\
$\text{Case study}$ & $F(1,55)=$ & $0,05$ & \\
$\text{Approach}$ $\times$ $\text{Case study}$ & $F(1,55)=$ & $1,78$ & \\
\bottomrule
\end{tabular}
\end{table}

Based on this findings we can reject the null hypothesis of mean times
spent to complete software development of selected cases are the same
for Object-oriented and Panorama approach. This leads to accepting
alternative hypothesis of mean times for development being significantly
different. Descriptive statistics in Table
\ref{tab:h1-descriptive-total-time} and graphical presentation of mean
time spent to complete software development of use cases clearly show
that time required to complete development is shorter with Panorama than
Object-oriented approach.

\begin{table}

\caption{\label{tab:h1-descriptive-total-time}Descriptive statistics of total time}
\centering
\begin{tabular}[t]{lrcr}
\toprule
Approach & Case study & N & Mean\\
\midrule
 & $C_1$ & $14$ & $12h \text{ } 59min$\\
\cmidrule{2-4}
 & $C_2$ & $14$ & $13h \text{ } 25min$\\
\cmidrule{2-4}
\multirow{-3}{*}{\raggedright\arraybackslash Object-oriented} & Total & $\boldsymbol{28}$ & $\boldsymbol{13h} \text{ } \boldsymbol{22min}$\\
\cmidrule{1-4}
 & $C_1$ & $14$ & $9h \text{ } 59min$\\
\cmidrule{2-4}
 & $C_2$ & $14$ & $9h \text{ } 40min$\\
\cmidrule{2-4}
\multirow{-3}{*}{\raggedright\arraybackslash Panorama} & Total & $\boldsymbol{28}$ & $\boldsymbol{9h} \text{ } \boldsymbol{49min}$\\
\cmidrule{1-4}
 & $C_1$ & $14$ & $11h \text{ } 29min$\\
\cmidrule{2-4}
 & $C_2$ & $14$ & $11h \text{ } 33min$\\
\cmidrule{2-4}
\multirow{-3}{*}{\raggedright\arraybackslash Total} & Total & $28$ & $11h \text{ } 31min$\\
\bottomrule
\end{tabular}
\end{table}

To conclude, there is a significant difference between the mean time
spent to complete software development of selected use cases using
Object-oriented and Panorama approach, with
\(\boldsymbol{F(1, 55) = 137,48}\) and level of significance
\(\boldsymbol{p < 0,05}\) and that time is shorter with Panorama than
object-oriented approach.

\begin{figure}

{\centering \includegraphics[width=0.9\linewidth]{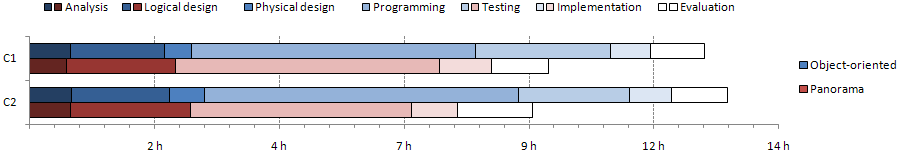} 

}

\caption{Mean time spent to complete software development of use cases}\label{fig:H1-mean-time}
\end{figure}

With use cases \(C_1\) and \(C_2\) the development with Panorama was
\(25\%\) faster than with Object-oriented approach, as depicted in
Figure \ref{fig:H1-mean-time}.

\paragraph{\texorpdfstring{Hypothesis
\(\text{H}_2\)}{Hypothesis \textbackslash{}text\{H\}\_2}}\label{hypothesis-texth_2}

Hypothesis \(H_2\) asserts that Panorama puts emphasis on analysis,
logical design and testing, while object-oriented approach on physical
design and programming.

When examining the data we can see from Figure \ref{fig:H1-mean-time}
that effort distribution by approaches (Object-oriented and Panorama) is
not equally divided. The comparison depicts distribution on Software
Development activities by approaches, where we can notice that in
Panorama approach majority of the time is spent in Analysis, Logical
design and Testing, while no time is allocated to Programming and
Physical design. In contrast to Panorama is in Object-oriented approach
time allocated to Programming and Physical design quite substantial. The
Programming and Physical design phase in Panorama is omitted, due to the
fact that user defines all the requirements in Analysis and Logical
design, while the code at implementation level is automatically
generated and supported by Panorama tool.

\begin{figure}

{\centering \includegraphics[width=1\linewidth]{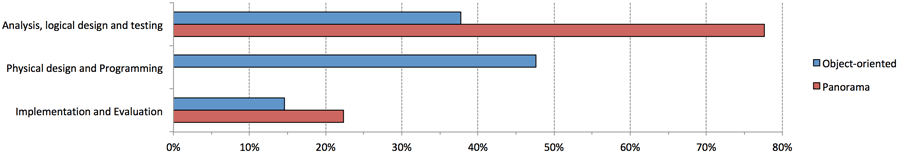} 

}

\caption{Mean effort distribution on Software Development activities grouped by significant factors}\label{fig:H2-mean-effort-distribution}
\end{figure}

For better comparison of effort distribution on Software Development
activities the visualization depicted in Figure
\ref{fig:H2-mean-effort-distribution} groups aforementioned phases and
approaches. The data is normalized regarding the total time required to
complete use cases \(C_1\) and \(C_2\) and belonging percentage of time
of selected phase is than displayed.

To test the significance of the results multiple \(1\)-way ANOVA tests
with factor \emph{Approach} and following groups of dependent variables:

\begin{itemize}
\tightlist
\item
  \emph{Analysis, logical design and testing};
\item
  \emph{Physical design and programming} and
\item
  \emph{Implementation and evaluation}
\end{itemize}

were conducted. Exploratory data analysis confirmed that data is
normally distributed and Levene's test also confirmed the assumption of
the ANOVA test. The results of ANOVA test and accompanying descriptive
statistics are depicted in, where we can confirm for all \(3\) factors
\(I\), \(II\) and \(III\) that the mean effort distribution between
Object-oriented and Panorama approach differs significantly (see Table
\ref{tab:h2-anova-and-descriptive}).

\begin{table}

\caption{\label{tab:h2-anova-and-descriptive}ANOVA results and descriptive statistics of combined effort distributions on Software Development activities}
\centering
\begin{tabular}[t]{rcrrc}
\toprule
Approach & N & Mean & F-value & Sig.\\
\midrule
\addlinespace[0.3em]
\multicolumn{5}{l}{\textbf{I) Analysis, logical design and testing}}\\
\hspace{1em}Object-oriented & $28$ & $\boldsymbol{37,76\%}$ &  & \\
\hspace{1em}Panorama & $28$ & $\boldsymbol{77,64\%}$ & $F(1,54) = \boldsymbol{1,755}$ & $\checkmark$\\
\hspace{1em}Total & $56$ & $57,70\%$ &  & \\
\addlinespace[0.3em]
\multicolumn{5}{l}{\textbf{II) Physical design and programming}}\\
\hspace{1em}Object-oriented & $28$ & $\boldsymbol{47,65\%}$ &  & \\
\hspace{1em}Panorama & $28$ & $\boldsymbol{0,00\%}$ & $F(1, 54) = \boldsymbol{4,668}$ & $\checkmark$\\
\hspace{1em}Total & $56$ & $23,83\%$ &  & \\
\addlinespace[0.3em]
\multicolumn{5}{l}{\textbf{III) Implementation and evaluation}}\\
\hspace{1em}Object-oriented & $28$ & $14,59\%$ &  & \\
\hspace{1em}Panorama & $28$ & $22,36\%$ & $F(1,54) = \boldsymbol{107,00}$ & $\checkmark$\\
\hspace{1em}Total & $56$ & $18,47\%$ &  & \\
\bottomrule
\end{tabular}
\end{table}

The most significant is difference in mean effort distributions of
\emph{Physical design} and \emph{programming} between Object-oriented
and Panorama approach with \(\boldsymbol{F(1, 54) = 4,668}\) and level
of significance \(\boldsymbol{p < 0,05}\), which confirms the fact that
in Panorama users don't spent time for physical design and programming,
because this is done automatically by underlying system. The second most
significant difference in mean effort distributions is of
\emph{Analysis, logical design and testing} between Object-oriented and
Panorama approach with \(\boldsymbol{F(1, 54) = 1,755}\) and level of
significance \(\boldsymbol{p < 0,05}\), which confirms the fact that
this phases are significantly more addressed in Panorama than in
Object-oriented approach. Mean effort distributions of
\emph{Implementation and evaluation} is with
\(\boldsymbol{F(1, 54) = 107,00}\) and level of significance
\(\boldsymbol{p < 0,05}\) also significant but is less notable than with
previous factors. Based on these findings we can confirm hypothesis
\(H_2\).

More detail interpretation of time allocations by software development
activities and critical overview can be found in
\citep{kruchten_what_2008}.

\paragraph{Questionnaire analysis}\label{questionnaire-analysis}

The results of the questionnaire are summarized in Table
\ref{tab:questionnaire-anova-and-descriptive} and grouped by selected
approach. Each of the \(7\) questions was rated on a Likert-type scale
from \(1\) to \(10\).

\begin{table}

\caption{\label{tab:questionnaire-anova-and-descriptive}ANOVA results and descriptive statistics of question analysis from questionnaire}
\centering
\begin{tabu} to \linewidth {>{\raggedleft}X>{\raggedright\arraybackslash}p{6.5cm}>{\centering}X>{\centering\arraybackslash}p{2cm}>{\centering}X>{\raggedleft\arraybackslash}p{3.5cm}>{\centering}X}
\toprule
\multicolumn{2}{c}{ } & \multicolumn{3}{c}{Mean score by approach} & \multicolumn{2}{c}{ } \\
\cmidrule(l{2pt}r{2pt}){3-5}
  &   & OO & Panorama & Total & F-value & Sig.\\
\midrule
C. & What is your experience with the methodology you are assessing? & $\boldsymbol{3,89}$ & $\boldsymbol{2,32}$ & $3,11$ & $F(1,54) = \boldsymbol{8,86}$ & $\checkmark$\\
D. & The notation is capable of expressing models of both static aspects of the system (e.g. structure) and dynamic aspects (e.g. processing)? & $\boldsymbol{8,57}$ & $\boldsymbol{4,32}$ & $6,45$ & $F(1,54) = \boldsymbol{187,15}$ & $\checkmark$\\
E. & Did you find the notation and modeling language easy to learn? & $\boldsymbol{6,07}$ & $\boldsymbol{7,57}$ & $6,82$ & $F(1,54) = \boldsymbol{38,04}$ & $\checkmark$\\
F. & Identify the required a priori technical knowledge for conducting analysis and design stage of software development life cycle! & $\boldsymbol{6,09}$ & $\boldsymbol{7,29}$ & $6,68$ & $F(1,54) = \boldsymbol{20,81}$ & $\checkmark$\\
G. & Estimate the effort required to progress from design into implementation stage! & $\boldsymbol{4,82}$ & $\boldsymbol{9,07}$ & $6,95$ & $F(1,54) = \boldsymbol{379,69}$ & $\checkmark$\\
\addlinespace
H. & How well is user participation encouraged and supported? & $\boldsymbol{7,04}$ & $\boldsymbol{9,32}$ & $8,18$ & $F(1,54) = \boldsymbol{87,63}$ & $\checkmark$\\
I. & Estimate the effort required introducing additional functionalities in working software product! & $\boldsymbol{4,46}$ & $\boldsymbol{8,21}$ & $6,34$ & $F(1,54) = \boldsymbol{267,93}$ & $\checkmark$\\
\bottomrule
\end{tabu}
\end{table}

The interpretation of Likert-scale values from Table
\ref{tab:questionnaire-anova-and-descriptive} are as follows:

\begin{enumerate}
\def\labelenumi{\Alph{enumi})}
\setcounter{enumi}{2}
\tightlist
\item
  What is your experience with the methodology you are assessing?

  \begin{itemize}
  \tightlist
  \item
    \(1\) - somewhat familiar, \(4\) - knows its details, \(7\) - I've
    used it, \(10\) - I've created it
  \end{itemize}
\item
  The notation is capable of expressing models of both static aspects of
  the system (e.g.~structure) and dynamic aspects (e.g.~processing)?

  \begin{itemize}
  \tightlist
  \item
    \(1\) - strongly disagree, \(4\) - disagree, \(7\) - agree, \(10\) -
    strongly agree
  \end{itemize}
\item
  Did you find the notation and modeling language easy to learn?

  \begin{itemize}
  \tightlist
  \item
    \(1\) - very hard, \(4\) - hard, \(7\) - fairly easy, \(10\) - very
    easy
  \end{itemize}
\item
  Identify the required a priori technical knowledge for conducting
  analysis and design stage of software development life cycle!

  \begin{itemize}
  \tightlist
  \item
    \(1\) - very detailed software development knowledge, \(3\) -
    advanced software development knowledge, \(7\) - basics of software
    development, \(10\) - intuitive approach, very little knowledge
    required
  \end{itemize}
\item
  Estimate the effort required to progress from design into
  implementation stage!

  \begin{itemize}
  \tightlist
  \item
    \(1\) - manually with user intervention, \(3\) - semi-automatically
    with major user interventions, \(7\) - semi-automatically with minor
    user interventions, \(10\) - automatically, without user
    intervention
  \end{itemize}
\item
  How well is user participation encouraged and supported?

  \begin{itemize}
  \tightlist
  \item
    \(1\) - very difficult, only professional developers are involved in
    software development, \(4\) - satisfactory, users are involved on
    their request, \(7\) - intermediate, with users involvement in early
    stages of software development, \(10\) - very well, the user can
    even build a simple system from scratch to working prototype
  \end{itemize}
\item
  Estimate the effort required introducing additional functionalities in
  working software product!

  \begin{itemize}
  \tightlist
  \item
    \(1\) - very demanding, it requires complete reorganization of
    software product and involvement of multiple actors, \(4\) -
    intermediate, new user requirements are passed to developers, \(8\)
    - minor changes can be done immediately with cooperation between
    users and developers, \(10\) - very intuitive, user can introduce
    changes to software product without the support of technical staff
  \end{itemize}
\end{enumerate}

The first question \((C)\) addressed evaluators' background and their
experience with given approach. As expected, evaluators had more
experience in Object-oriented approach (mean = \(3,89\)) than Panorama
(mean = \(2,32\)). That's because they previously hadn't worked with
Panorama but evaluators \(S_1\) to \(S_{24}\) had introductory knowledge
and \(P_1\) to \(P_4\) had experience in Object-oriented approach. Mean
experience with the methodology is with \(\boldsymbol{F(1, 54) = 8,86}\)
and level of significance \(\boldsymbol{p < 0,05}\), which confirms that
evaluators have different experience in selected approaches.

\paragraph{\texorpdfstring{Hypothesis
\(\text{H}_3\)}{Hypothesis \textbackslash{}text\{H\}\_3}}\label{hypothesis-texth_3}

The claim of \(H_3\) is that Panorama is more appropriate for domains
with static (data oriented) rather than dynamic (process oriented)
components. This aspect of development approach was addressed in
question \(D\) (see Table \ref{tab:questionnaire-anova-and-descriptive})
where users had to indicate whether the approach is capable of
expressing models of both static aspects of the system (e.g.~structure)
and dynamic aspects (e.g.~processing). The difference in \emph{mean
notation capability of expressing dynamics} with
\(\boldsymbol{F(1, 54) = 187,15}\) and level of significance
\(\boldsymbol{p < 0,05}\), which confirms that null hypothesis can be
rejected and alternative hypothesis is confirmed. Mean notation
capability of expressing dynamics in Object-oriented approach is
\(\boldsymbol{8,57}\) and with Panorama \(\boldsymbol{4,32}\), that
leads to conclusion that Panorama with score of \(4,32\) on a \(1\) to
\(10\) scale is not appropriate for problem domains with emphasis on
processing but rather for data oriented applications. Object-oriented
approach has, as expected, scored very high score, because its notation
is capable of expressing statics and dynamics of the system.

\paragraph{\texorpdfstring{Hypothesis
\(\text{H}_4\)}{Hypothesis \textbackslash{}text\{H\}\_4}}\label{hypothesis-texth_4}

Hypothesis \(H_4\) asserts that with Panorama approach in contrast to
Object-oriented, users are more encouraged to participate, required
technical knowledge for development is lower, introducing additional
functionalities is less demanding and the approach is easier to learn.
The aforementioned hypothesis is concerned about questions \(E\), \(F\),
\(G\), \(H\) and \(I\) from the questionnaire depicted in Table
\ref{tab:questionnaire-anova-and-descriptive}.

Based on analysis of aforementioned questions from Table
\ref{tab:questionnaire-anova-and-descriptive} (simplicity to learn the
approach, low a priori technical knowledge, simplicity of progress from
design to implementation, user participation and simplicity of
introducing additional functionalities) we can confirm with significance
that hypothesis \(H_4\) is valid.

\paragraph{Association and bias of
raters}\label{association-and-bias-of-raters}

\textbf{Intraclass correlation (ICC)} across raters \(P_1\) to \(P_4\)
and \(S_1\) to \(S_{24}\) was analyzed. In our experiment this was the
case \(2\) ICC with the same set of raters that rate each subject, which
corresponds to a fully-crossed \((rater \times subject)\), \(2\)-way
ANOVA design in which both subject and rater are separate effects. There
were \(28\) raters (\(P_1\) to \(P_4\) and \(S_1\) to \(S_{24}\)) and
\(30\) subjects (\(8\) questions about the development process and \(7\)
effort distributions of development time) included in ICC calculation.
In case \(2\), rater is considered a random affects, which means the
raters in the study are considered a random sample from a population of
potential raters. The case \(2\) ICC estimates the reliability of the
larger population of raters. ICC was used separately to evaluate expert
and student users and to estimate the reliability of a single rating
(consistency) and the reliability of a mean of several ratings (absolute
agreement). The result of both measurements is depicted in Table
\ref{tab:ICC}.

\begin{table}

\caption{\label{tab:ICC}ANOVA results and descriptive statistics of question analysis from questionnaire}
\centering
\begin{tabu} to \linewidth {>{\raggedright\arraybackslash}p{3cm}>{\centering}X>{\centering}X>{\centering}X>{\centering}X}
\toprule
\multicolumn{1}{c}{ } & \multicolumn{2}{c}{Consistency} & \multicolumn{2}{c}{Absolute agreement} \\
\cmidrule(l{2pt}r{2pt}){2-3} \cmidrule(l{2pt}r{2pt}){4-5}
  & Single measures $\boldsymbol{ICC(2,1)}$ & Average measures $\boldsymbol{ICC(2,k)}$ & Single measures $\boldsymbol{ICC(2,1)}$ & Average measures $\boldsymbol{ICC(2,k)}$\\
\midrule
$\textbf{Experts } (k=4)$ & $0,949$ & $0,987$ & $0,943$ & $0,985$\\
$\textbf{Students } (k=24)$ & $0,959$ & $0,998$ & $0,959$ & $0,998$\\
$\textbf{All } (k=28)$ & $0,948$ & $0,998$ & $0,946$ & $0,998$\\
\bottomrule
\end{tabu}
\end{table}

The results in Table \ref{tab:ICC} denote that there exist a
statistically significant positive correlation between all raters -
experts and students. Both reliability of a single rating and
reliability of a mean of several ratings are high. ICC approaches \(1\)
when there is no variance within subjects, indicating total variation in
measurements on the Likert scale is due solely to the target variable.
ICC is high because any given subject tends to have the same score
across the raters. ICC is also interpreted as the ratio of variance
explained by the independent variable divided by total variance, where
total variance is the explained variance plus variance due to the raters
plus residual variance. High ICC is a result of no or little variance
due to the raters and no residual variance to explain.

Based on the nature of Case \(2\) intraclass correlation with random
effects that was used, the results obtained can be generalized to other
raters.

\subsubsection{Discussion}\label{discussion-1}

The results show that time spent to complete software development of
selected use cases \(C_1\) and \(C_2\) is shorter when using Panorama
approach opposed to using Object-oriented approach (see Figure
\ref{fig:H1-mean-time}). In Panorama approach the emphasis is on logical
design stage, where business vocabulary is constructed and on testing
stage, where actual data is inserted into the system and actual testing
is performed. On the other hand with Object-oriented approach extra work
is conducted in programming and physical design stage.

In average this results in \(25\%\) less time required to complete
software development life cycle using Panorama approach towards
Object-oriented. It has to be noted that this is valid for problem
domains with emphasis on data oriented aspects rather than process
oriented.

The results of questionnaire analysis (see Table
\ref{tab:h2-anova-and-descriptive}) show that the experience of raters
is higher in Object-oriented approach due to raters' background and high
dissemination of Object-oriented approach in modern software development
projects. Panorama approach is not as strong as Object-oriented approach
in expressing dynamic aspects (e.g.~processing) of the system, while it
focuses on static aspects (e.g.~structure). Notation and modeling
language is slightly easier to learn in Panorama approach as it is
mainly focused on data modeling techniques. The same applies to a priori
technical knowledge for analysis and design. As progressing from design
to implementation stage is concerned Panorama turns out to be preferred
approach due to availability of automatic code generation and
transformation from logical to physical design. Following JAD approach
Panorama also emphasizes user participation and received better score
than Object-oriented approach. An important aspect is also the effort
for introducing additional functionalities in working software product,
whereas Panorama approach scored higher than object-oriented approach
because of almost nonexistent programming phase which is automatically
implemented in Panorama tool.

\section{Conclusions and future work}\label{conclusion-and-future-work}

In this paper a software solution for facilitating associative thinking
paradigm, Panorama, was presented. The main goal was to create a ``user
friendly'' tool for information management and software development for
people with low IT skills. This was accomplished by introducing
visualization in multi-dimensional space with focus on the chosen node
(object of observation) and its nearest neighbours (object's context).
Only semantically nearest nodes with appropriate attributes are being
displayed with a possibility of freely traversing through information
space by changing to the new object of observation. The user interface
we have developed is very simple and pursues presenting and organizing
the information on the screen in a way that user is not overloaded but
still has the ability to expand the context and view the situation in
all its extent (following focus + context information visualization
techniques).

We can conclude that the conceptual idea of using associative thinking
paradigm in computer aided software development was successfully
realized in several real world cases. By following the proposed
methodology with the Panorama tool, the user is given an opportunity to
design the application and start using it, without extensive technical
knowledge. Panorama especially minimizes the impact on the last two
phases in Software Development Life Cycle (implementation and review),
due to existence of all required mechanisms for updating and viewing the
content of data vocabulary and controlling the input of data with
referential integrity checking. Future trends in information
visualization were also addressed. One of them was utilization of
information visualization techniques in information systems development
process that was realized in Panorama tool.

With the detailed comparison of Panorama approach to existing software
development approaches Panorama is positioned as a specific
problem-solving approach. It is not intended for process oriented
domains, but rather for smaller environments, especially office
management, CRM, record management, planning, maintenance etc. The
conducted experiment pointed out that there was less time spent to
complete software development of selected use cases than using
Object-oriented techniques and approaches. One of the findings is also
the effort distribution of Software Development activities. In
Object-oriented approaches emphasis is on design and programming, while
Panorama eliminates programming activity and is focused mainly on design
and testing.

This research has several limitations as already pointed out in section
\ref{evaluation}. The Panorama approach is suitable only for problem
domains that emphasis data oriented problem domains while it lacks to
support the ones with process oriented aspects.

The results of this research could also assist ontology management
community, where Panorama can be used as an ontology editor. Export of
Panorama's data vocabulary and known objects space to XML format is
already supported, but clearly further work is required to make data
available in ontological languages with more expressive power. In our
future work we will therefore expand the variety of export formats that
Panorama supports. Data from Panorama available in a form of ontology
could enable reuse of knowledge in intelligent systems with inference
capabilities to derive new knowledge.

\end{document}